\documentclass{emulateapj}

\usepackage{graphicx,textcomp,longtable}
\usepackage{natbib}

\newcommand{\kms}{km s$^{-1}$}
\newcommand{\ms}{m s$^{-1}$}
\newcommand{\vsini}{v\sin{i_\star}}
\newcommand{\Teff}{$T_{\text{eff}}$}
\newcommand{\bjdtdb}{\ensuremath{\rm {BJD_{TDB}}}}
\newcommand{\hjdutc}{\ensuremath{\rm {HJD_{UTC}}}}

\defcitealias{Albrecht12a}{Albrecht et al. 2012a}
\defcitealias{Albrecht12}{Albrecht et al. 2012b}

\shorttitle{Obliquity of HAT-P-17}
\shortauthors{Fulton et al.}

\begin{document}

%%%%%%%%%%%%%%%%%%%%%%%%%%%%%%%%%%%%%%%%%%%%%%%%%%%%%
\title{The stellar obliquity and the long-period planet in the HAT-P-17 exoplanetary system}

%%%%%%%%%%%%%%%%%%%%%%%%%%%%%%%%%%%%%%%%%%%%%%%%%%%%%
% author list
\author{
Benjamin J.~Fulton\altaffilmark{1},
Andrew W.~Howard\altaffilmark{1},
Joshua N.~Winn\altaffilmark{2}, 
Simon Albrecht\altaffilmark{2},
Geoffrey W.~Marcy\altaffilmark{4},
Justin R.~Crepp\altaffilmark{6},
Gaspar A.~Bakos\altaffilmark{5,7,8},
John Asher Johnson\altaffilmark{3,7},
Joel D.~Hartman\altaffilmark{5},
Howard Isaacson\altaffilmark{4},
Heather A.~Knutson\altaffilmark{3},
Ming Zhao\altaffilmark{9}
}
%%%%%%%%%%%%%%%%%%%%%%%%%%%%%%%%%%%%%%%%%%%%%%%%%%%%%
% affiliations

% Hawaii
\altaffiltext{1}{Institute for Astronomy, University of Hawaii at Manoa, 2680 Woodlawn Dr, Honolulu, HI 96822}
%  MIT/Kavli 
\altaffiltext{2}{Department of Physics and Kavli Institute for Astrophysics and Space Research, Massachusetts Institute of Technology, Cambridge, MA 02139, USA}
% Caltech
\altaffiltext{3}{California Institute of Technology, Pasadena, CA, USA}
% UCB
\altaffiltext{4}{University of California, Berkeley, Berkeley, CA, USA}
% CfA
\altaffiltext{5}{Princeton University, Department of Astrophysical Sciences, Princeton, NJ, USA}
%Notre Dame
\altaffiltext{6}{University of Notre Dame, Department of Physics, Notre Dame, IN, USA}
%225 Nieuwland Science Hall

%Other
\altaffiltext{7}{Alfred P. Sloan Fellow}
\altaffiltext{8}{Packard Fellow}

\altaffiltext{9}{Department of Astronomy and Astrophysics, 525 Davey Laboratory, The Pennsylvania State University, University Park, PA 16802, USA}

%%%%%%%%%%%%%%%%%%%%%%%%%%%%%%%%%%%%%%%%%%%%%%%%%%%%%
\begin{abstract}
We present the measured projected obliquity --~the sky-projected angle between the stellar spin axis and orbital angular momentum --~of the inner planet of the HAT-P-17 multi-planet system. We measure the sky-projected obliquity of the star to be $\lambda=19^{+14}_{-16}$ degrees by modeling the Rossiter-McLaughlin (RM) effect in Keck/HIRES radial velocities (RVs). The anomalous RV time series shows an asymmetry relative to the midtransit time, ordinarily suggesting a nonzero obliquity --~but in this case at least part of the asymmetry may be due to the convective blueshift, increasing the uncertainty in the determination
of $\lambda$. We employ the semi-analytical approach of \citet{Hirano11} that includes the effects of macroturbulence, instrumental broadening, and convective blueshift to accurately model the anomaly in the net RV caused by the planet eclipsing part of the rotating star. Obliquity measurements are an important tool for testing theories of planet formation and migration. To date, the measured obliquities of $\sim$50 Jovian planets span the full range, from prograde to retrograde, with planets orbiting cool stars preferentially showing alignment of stellar spins and planetary orbits. Our results are consistent with this pattern emerging from tidal interactions in the convective envelopes of cool stars and close-in planets.
In addition, our 1.8 years of new RVs for this system show that the orbit of the outer planet is more poorly constrained than previously thought, with an orbital period now in the range of 10--36 years.
\end{abstract}
%%%%%%%%%%%%%%%%%%%%%%%%%%%%%%%%%%%%%%%%%%%%%%%%%%%%%

%%%%%%%%%%%%%%%%%%%%%%%%%%%%%%%%%%%%%%%%%%%%%%%%%%%%%
\section{Introduction}
HAT-P-17 is an early K dwarf star that hosts a transiting Saturn-mass planet (planet b) on a 10.3 day orbit and a more massive outer companion (planet c) on a long-period orbit \citep[][hereafter H12]{Howard12}. Transits of planet b were discovered in 2010 by the Hungarian-made Automated Telescope Network \citep[HATNet,][]{Bakos04}. Followup Keck/HIRES RVs were used to measure the mass of planet b and enabled the discovery of planet c. More than 150 hot Jupiters have been discovered, but it appears that hot Jupiters tend to lack additional short-period giant planet companions \citep{Steffen12}. HAT-P-17 is one of only six of systems with a transiting Jovian-sized planet and an additional substellar companion. The five other systems include HAT-P-13 \citep{Bakos09}, HAT-P-31 \citep{Kipping11}, Kepler-9 \citep{Holman10}, Kepler-30 \citep{Fabrycky12}, and KOI-94 \citep{Hirano12}. These rare multi-planet transiting Jovian systems provide important insight into the formation and evolution of hot Jupiters.

Current theory suggests that Jovian planets form at orbital distances of $\gtrsim$1 AU where additional protoplanetary solids (ice) augment their formation. They then migrate inwards to become hot Jupiters. Popular theories that attempt to explain their resulting close-in orbits involve a 3rd body (in addition to the Jovian planet and it's host star) that perturbs the orbit of the soon-to-be hot Jupiter and excites high eccentricities through either the Kozai mechanism or planet-planet scattering. This highly eccentric orbit then decays through tidal interactions into a close-in circular orbit \citep{Nagasawa08, Fabrycky07, Naoz11}. This scenario would produce hot Jupiters with a large range of orbital obliquities. Others suggest that hot Jupiters migrate within the circumstellar disk from which they formed through interactions with the disk \citep{Lin96}. In this case we expect that all of the bodies would lie in coplanar orbits that are all well-aligned with the stellar spin axis.
If the orbit of planet b is aligned to host star's spin, it would suggest that this system was formed by migration rather than perturbation if the two planets are coplanar.
%Although we cannot infer the orbital inclination of planet c, if it could be shown that the two planets are coplanar planet b's low orbital obliquity may indicate that this system was formed by migration rather than perturbation.
A coplanar and apsidally locked geometry would also allow for a precise measurement of the interior density structure of planet b \citep{Batygin09,Mardling10}.

An emerging trend suggests that hot Jupiters around cool stars ($T_{\text{eff}} \lesssim 6250$ K) with large convective envelopes tend to be better aligned with their host star's rotation axis \citep{Albrecht12}. Tidal energy is most efficiently dissipated by turbulent eddies in the convective regions of stars \citep{Zahn08}. As a result, the rate of tidal dissipation depends on the mass of the convective envelope. Strong tidal interactions with the convective envelope force the system into alignment in a relatively short time. Stars hotter than 6250 K have small or no convective envelopes, and it takes much longer for the system to align \citep{Winn10a}. HAT-P-17 is a cool star with \Teff\ $\sim5200$ K, but planet b's orbital distance is relatively large making tidal interactions weak. According to the tidal figure of merit devised by \citet{Albrecht12}, we would expect the tidal dissipation rate for this system to be too slow to cause obliquity damping, despite the star's thick convective envelope. This makes HAT-P-17 an interesting test case.
%Given the estimated age of the star (7.8$\pm 3.3$ Gyr, H12) we do not expect the system to be aligned.

%We measure the sky-projected orbital obliquity ($\lambda$) of the HAT-P-17 multi-planet system using a semi-analytical model \citep{Hirano11} that realistically describes instrumental and physical sources of line broadening that affect the measured radially velocity anomaly (RM effect) during a spectroscopic transit. We also find that the convective blueshift \citep{Shporer11} contributes significantly to the shape of the RM effect and must be included in our model for an accurate measurement of $\lambda$. This model, combined with Differential-Evolution Markov Chain Monte Carlo \citep[DE-MCMC,][]{Braak06} modeling allowed us to quickly and accurately estimate the parameters and their associated errors.

In this work we revisit the orbital parameters of planet c with new Keck/HIRES RV and Keck/NIRC2 adaptive optics images, and present a measurement of the sky-projected orbital obliquity of the star relative to planet b. In \S \ref{obs} we discuss our observational techniques. We discuss our RV and RM modeling and results in \S \ref{analysis}, and in \S \ref{discussion} we interpret and summarize our findings.

%%%%%%%%%%%%%%%%%%%%%%%%%%%%%%%%%%%%%%%%%%%%%%%%%%%%%

%%%%%%%%%%%%%%%%%%%%%%%%%%%%%%%%%%%%%%%%%%%%%%%%%%%%%
\section{Observations}
\label{obs}
%%%%%%%%%%%%%%%%%%%%%%%%%%%%%%%%%%%%%%%%%%%%%%%%%%%%%

\subsection{Keck/HIRES Spectroscopy}

Since the publication of \citet{Howard12}, we 
have measured the RV of HAT-P-17 (V=10.54) for an additional 
1.8 years using HIRES \citep{Vogt94} on the Keck I telescope.  
We adopted the same observing strategy and Doppler analysis techniques 
described in Section 2.3 of H12.  
In brief, we observed HAT-P-17 through a cell of gaseous iodine and 
measured the subtle Doppler shifts of the stellar lines with respect 
to the reference iodine lines using a forward modeling analysis \citep{Butler96}.  

Our observations were designed to measure the Keplerian orbits of 
HAT-P-17b and c and also to measure the obliquity of HAT-P-17.  
For the latter, we observed a transit of HAT-P-17b on UT 26 August 2012.  
Our observing sequence lasted nearly six hours and bracketed the 3.2 hour long transit.   
We made 42 observations of $\sim$500 second duration separated by 45 second detector reads. 
To constrain the Keplerian slope, 
we made three additional observations on the same night approximately 3.8 hours 
after transit egress.
%(I didn't realize this. I'm not using these in my RM fit, should I be? They are included in the Keplerian fit)

Julian dates of the photon-weighted exposure mid-times were recorded during the observations, and then later converted to \bjdtdb\ using the tools described in \citet{Eastman10}\footnote{IDL tools for time systems conversion; http://astroutils.astronomy.ohio-state.edu/time/}. The photon-weighted exposure times are only accurate to $\sim$1 second due to internal limitations
%(what really causes this? finite sampling rate?)
of the exposure meter.

The complete set of RV measurements and their uncertainties are listed in Table \ref{tab:data}. 
These 100 RVs include 42 RVs from \citet{Howard12}, 
45 new RVs taken on the night of UT 26 August 2012 to measure the 
RM effect, and 13 additional RVs taken sporadically 
between 2010 April and 2012 December to measure the orbit of HAT-P-17c.

\tablenum{1}
\begin{deluxetable}{lrr}
\label{tab:data}
%\tablewidth{0pt} 
%\tabletypesize{\small}
\tablecaption{Radial velocity data\tablenotemark{a}}
\tablehead{\colhead{Time} & \colhead{RV} & \colhead{$\sigma_{\text{RV}}$}\\
\colhead{BJD$_{\text{TDB}}-2440000$} & \colhead{\ms} & \colhead{\ms}}

\startdata
14396.8272772 & -5.25 & 1.62 \\
14397.7946382 & -32.79 & 1.60 \\
14427.7815123 & -10.63 & 1.58 \\
14429.8199962 & -68.92 & 1.78 \\
14430.8485952 & -97.84 & 1.90 \\
14454.7162454 & 9.70 & 2.66 \\
14455.7078194 & 14.89 & 1.90
\enddata

\tablenotetext{a}{This table is available in its entirety in machine-readable form in the online journal. A portion is shown here for guidance regarding its form and content.}
\end{deluxetable}

\subsection{KECK/NIRC2 adaptive optics imaging}

In order to search for additional companions and sources of possible photometric dilution, we obtained high spatial resolution images of HAT-P-17 using NIRC2 (instrument PI: Keith Matthews) at the Keck II telescope on 2012-05-07 UT. Photometric dilution would affect the radius of HAT-P-17b measured by H12, and the presence of a physically associated companion would put constraints on our radial velocity fit. In addition, a statistical sample of the wide companions to exoplanet host stars may help our understanding of planetary formation mechanisms. Our observations consist of dithered images acquired using the KÕ filter (central wavelength = $2.12 \mu$m). We used the small camera setting to provide fine spatial sampling of the instrument point spread function.  The total on-source integration time was 16.2 seconds. Images were processed using standard techniques to flat-field the array, replace hot pixels, subtract the thermal background, and align and co-add individual frames.

%%%%%%%%%%%%%%%%%%%%%%%%%%%%%%%%%%%%%%%%%%%%%%%%%%%%%
\section{Analysis}
\label{analysis}
%%%%%%%%%%%%%%%%%%%%%%%%%%%%%%%%%%%%%%%%%%%%%%%%%%%%%
\subsection{Radial velocities}
\label{RV}

\begin{figure*}[ht]
\epsscale{1.0}
\plotone{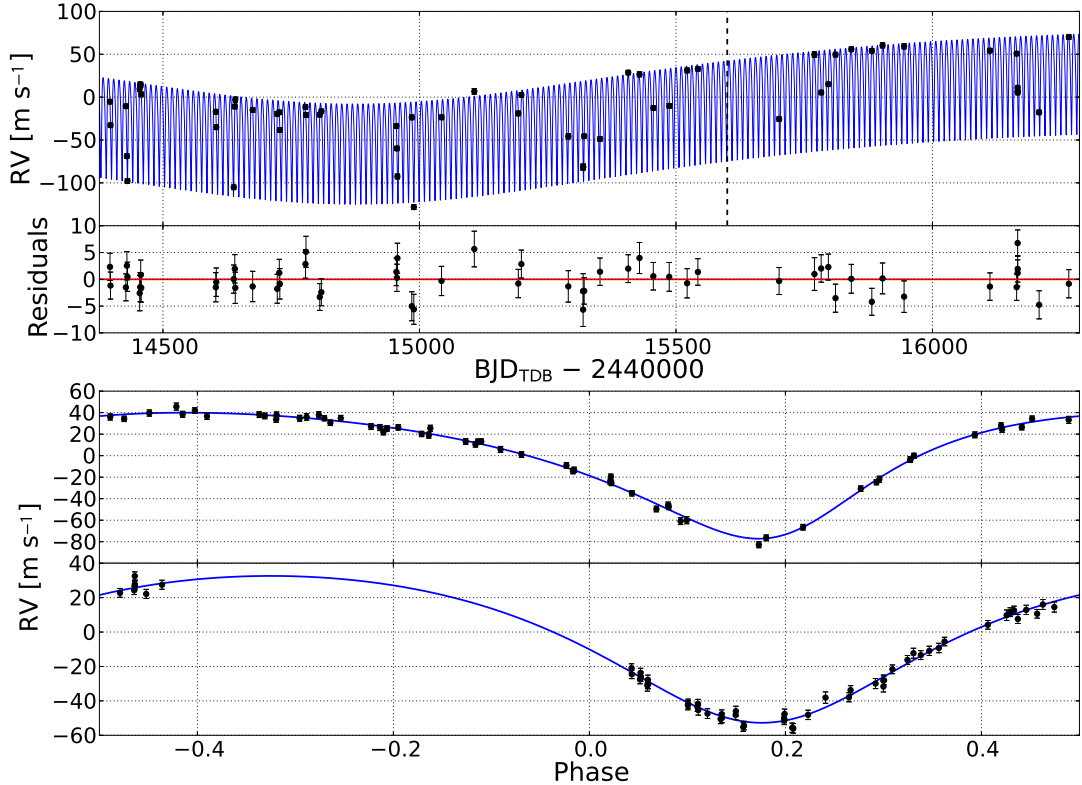}
\centering
\caption{
\emph{Top:} Keck/HIRES RV measurements for HAT-P-17 as a function of \bjdtdb\ with the best-fitting two-planet model and associated residuals found by $\chi^{2}$ minimization of the DE-MCMC chains. A stellar ``jitter" term (see Table \ref{tab:rv}) has been added in quadrature to the measurement errors. Data taken during the transit of HAT-P-17b for the purpose of measuring the RM effect was excluded from the RV fit and are not included in this plot. Data to the right of the vertical dashed line are new to this work, and data to the left are from H12. \emph{Bottom:} same RV measurements phase-folded to the orbital ephemerides of planets b (upper) and c (lower). Phase 0 corresponds to the time of mid-transit (or hypothetical transit). In each case the orbit of the other planet and an arbitrary center of mass velocity relative to a template spectrum ($\gamma$) has been removed.}
\label{rv}
\end{figure*}

\begin{figure}[h]
\epsscale{1.2}
\plotone{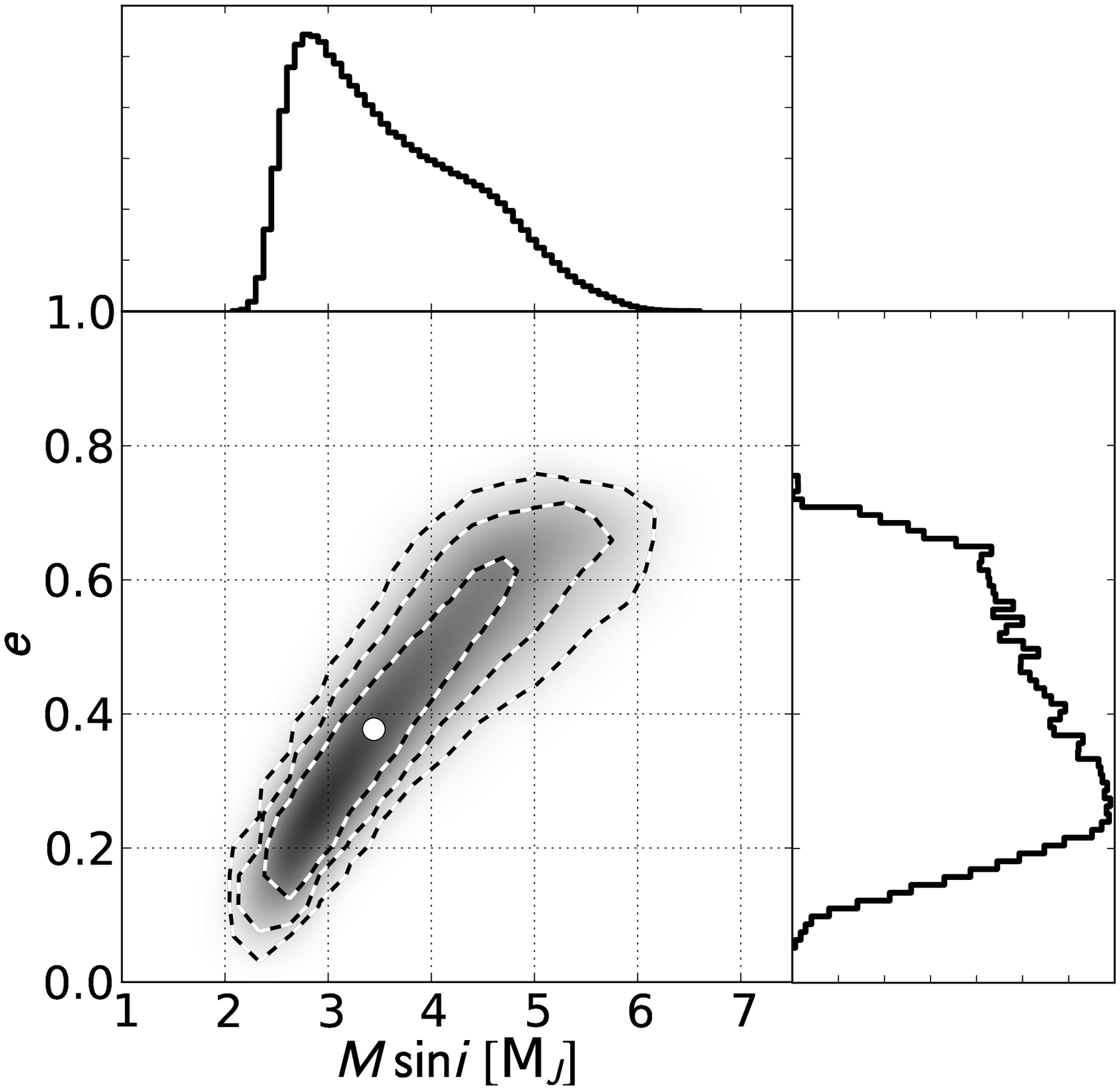}
\centering
\caption{
Probability distribution of eccentricity vs. minimum mass ($M\sin{i}$) in Jupiter masses for planet c from the DE-MCMC analysis. The dashed lines are 68\%, 95\%, and 99\% confidence intervals and the white dot is the median value. The median value is offset from the mode (most likely value) due to the asymmetric posterior distributions.}
\label{emsini}
\end{figure}

With 1.8 years of new radial velocities we revisited the orbital parameters of the outer companion in the HAT-P-17 system (planet c). We analyzed the entire dataset with a custom version of EXOFAST\footnote{IDL code available at; http://astroutils.astronomy.ohio-state.edu/exofast/} \citep{Eastman12} ported to Python (ExoPy hereafter). ExoPy utilizes the Differential-Evolution Markov Chain Monte Carlo (DE-MCMC) technique \citep{Braak06} to find the best-fitting parameters and their associated uncertainties. We also ported a subset of the RVLIN\footnote{IDL code available at; http://exoplanets.org/code/} \citep{Wright09} package to Python for quick calculation of the Keplerian orbit model. We fit a 12 parameter model to the radial velocity data that included the period ($P_{b}$), time of transit ($T_{\rm tra,b}$), eccentricity ($e_{b}$), argument of periastron of the star's orbit ($\omega_{b}$), radial velocity semi-amplitude ($K_{b}$) of planet b, the same parameters for planet c ($P_{c}$, $T_{\rm conj,c}$, $e_{c}$, $\omega_{c}$, $K_{c}$), the center of mass velocity of the system normalized to an arbitrary reference spectrum ($\gamma$), and a stellar ``jitter" term.
%, and an overall slope when needed ($\dot{\gamma}$).

We computed 24 DE-MCMC chains in parallel, continuously checking for convergence following the prescription of \citet{Eastman12}. We considered the chains well-mixed and halted the DE-MCMC run when the number of independent draws \citep[$T_{z}$, as defined in][]{Ford06} was greater than 1000 and the Gelman-Rubin statistic \citep{Gelman03,Ford06,Holman06} was within 1\% of unity for all parameters. In order to speed convergence, ensure that all parameter space was adequately explored, and minimize biases in parameters that physically must be finite and positive, we step in the modified and/or combinations of parameters shown in Table \ref{tab:rv}. Namely, due to the highly correlated uncertainties of $e$ with $\omega$ and $\vsini$ with $\lambda$ we step in $\sqrt{e}\cos{\omega}$, $\sqrt{e}\sin{\omega}$, $\sqrt{\vsini}\cos{\lambda}$, and $\sqrt{\vsini}\sin{\lambda}$ \citep{Eastman12,Albrecht12}.

We assigned Gaussian priors to $P_{b}$ and $T_{\rm tra}$ from the values given in the HAT-P-17b,c discovery paper ($P_{b} = 10.338523 \pm 9\times10^{-6}$ days, $T_{\rm tra,b} = 2454801.16943 \pm 2\times10^{-4}$ \bjdtdb, H12) that came from the highly constraining photometric transit data, and we assigned uniform priors to all other step parameters. We ignored any variation in transit times due to perturbations caused by planet c, but these are expected to be at least an order of magnitude smaller than the propagated uncertainty on $T_{\rm tra,b}$ (H12). We excluded the spectroscopic transit data from the Keplarian radial velocity fit because the higher density of data on the one night could bias the results in the presence of short-term ($\sim$hours) trends. The best fitting values and upper and lower ``1$\sigma$" errors for each parameter were determined by taking the median, 85.1, and 15.9 percentile values, respectively, of the resulting posterior distributions.

The results of the RV analysis are presented in Table \ref{tab:rv}. All of the parameters for planet b are consistent with the values from H12. However, with the new RV data we can now see that the period of planet c is much longer than the initially reported period of 1620 $\pm$ 20 days. Our RV time-series span a total timespan of 1869 days. We still have not seen a complete orbit of planet c, and thus the fit is quite poorly constrained. Figure \ref{emsini} shows the probability distributions of $M_{c}\sin{i_{c}}$ and eccentricity, and indicates that the allowed mass range ($3 \sigma$) for planet c is 2-6 $M_{J}$.

We also explored the possibility of a 4th body causing a linear trend ($\dot{\gamma}$) in the radial velocities in addition to the signal from planet c. Fits that included $\dot{\gamma}$ as a free parameter preferred a slope consistent with zero ($\dot{\gamma}=-3.7^{+5.5}_{-8.0}$ \ms yr$^{-1}$) with a 3 $\sigma$ limit of $|\dot{\gamma}| \leq 19$ \ms yr$^{-1}$. To assess the validity of adding one more free parameter to our model we calculate the bayesian information criterion (BIC):
\begin{equation}
\label{BIC}
\mbox{BIC} \equiv \chi^{2} + k\ln{n} 
\end{equation}
where $k$ is the number of degrees of freedom, and $n$ is the number of data points in the fit \citep{Liddle07}. The BIC increased when $\dot{\gamma}$ was a free parameter (104 vs. 100 with or without $\dot{\gamma}$ as a free parameter respectively). The BIC increase, our model fit favoring $\dot{\gamma}=0.0$, and the AO image (see \S \ref{sec:AO}) all indicate that the data are better described by a model with $\dot{\gamma}$ fixed at zero.

\tablenum{2}
\begin{deluxetable}{lrr}[h]
\label{tab:rv}
\tablewidth{0pt} 
\tabletypesize{\small}
\tablecaption{Radial velocity MCMC results}
\tablehead{\colhead{Parameter} & \colhead{Value} & \colhead{Units}}

\startdata
\sidehead{\bf {RV Step Parameters:}}
log($P_{b}$) & 1.0144585 $\pm 3.7e-07$ & log(days)\\
$T_{\rm tra,b}$ & 2454801.1702 $\pm 0.0003$ & \bjdtdb\\
$\sqrt{e_{b}}\cos{\omega_{b}}$ & -0.5442 $^{+0.0052}_{-0.0051}$ & \\
$\sqrt{e_{b}}\sin{\omega_{b}}$ & -0.214 $\pm 0.016$ & \\
log($K_{b}$) & 1.7678 $\pm 0.0051$ & \ms\\
log($P_{c}$) & 3.75 $^{+0.38}_{-0.2}$ & log(days)\\
$T_{\rm conj,c}$ & 2454146 $^{+100}_{-170}$ & \bjdtdb\\
$\sqrt{e_{c}}\cos{\omega_{c}}$ & -0.63 $^{+0.15}_{-0.16}$ & \\
$\sqrt{e_{c}}\sin{\omega_{c}}$ & -0.017 $^{+0.068}_{-0.061}$ & \\
log($K_{c}$) & 1.689 $^{+0.08}_{-0.061}$ & \ms\\
$\gamma$ & 20 $^{+27}_{-16}$ & \ms\\
$\dot{\gamma}$ & $\equiv$ 0.0  & \ms day$^{-1}$\\
$\ddot{\gamma}$ & $\equiv$ 0.0  & \ms day$^{-2}$\\
log(jitter) & 0.312 $^{+0.076}_{-0.081}$ & log(\ms)\\
\\
\hline
\sidehead{\bf {RV Model Parameters:}}
$P_{b}$ & 10.338523 $^{+8.8e-06}_{-8.9e-06}$ & days\\
$T_{\rm tra,b}$ & 2454801.1702 $\pm 0.0003$ & \bjdtdb\\
$T_{\rm peri,b}$ & 2454803.24 $\pm 0.05$ & \bjdtdb\\
$e_{b}$ & 0.3422 $\pm 0.0046$ & \\
$\omega_{b}$ & 201.5 $\pm 1.6$ & degrees\\
$K_{b}$ & 58.58 $^{+0.69}_{-0.68}$ & \ms\\
$P_{c}$ & 5584 $^{+7700}_{-2100}$ & days\\
$T_{\rm conj,c}\tablenotemark{$\dagger$}$ & 2454146 $^{+100}_{-170}$ & \bjdtdb\\
$T_{\rm peri,c}$ & 2454885 $^{+45}_{-57}$ & \bjdtdb\\
$e_{c}$ & 0.39 $^{+0.23}_{-0.17}$ & \\
$\omega_{c}$ & 181.5 $^{+5.3}_{-6.7}$ & degrees\\
$K_{c}$ & 48.8 $^{+9.9}_{-6.4}$ & \ms\\
$\gamma$ & 20 $^{+27}_{-16}$ & \ms\\
$\dot{\gamma}$ & $\equiv$ 0.0  & \ms day$^{-1}$\\
$\ddot{\gamma}$ & $\equiv$ 0.0  & \ms day$^{-2}$\\
jitter & 2.05 $^{+0.39}_{-0.35}$ & \ms\\
\\
\hline
\sidehead{\bf {RV Derived Parameters:}}
%$M_{b}\sin{i_{b}}$ & 0.532 $^{+0.018}_{-0.017}$ & $M_{J}$ \\ 
$M_{b}$ & 0.532 $^{+0.018}_{-0.017}$ & $M_{J}$ \\ 
$a_{b}$ & 0.0882 $\pm 0.0014$ & AU \\
$M_{c}\sin{i_{c}}$ & 3.4 $^{+1.1}_{-0.7}$ & $M_{J}$ \\
$a_{c}$ & 5.6 $^{+3.5}_{-1.4}$ & AU \\
\enddata

\tablenotetext{$\dagger$}{Since planet c is not known to transit, $T_{\rm conj,c}$ refers to the time of inferior conjunction.}
\end{deluxetable}

\begin{figure*}[ht]
\epsscale{1.1}
\plottwo{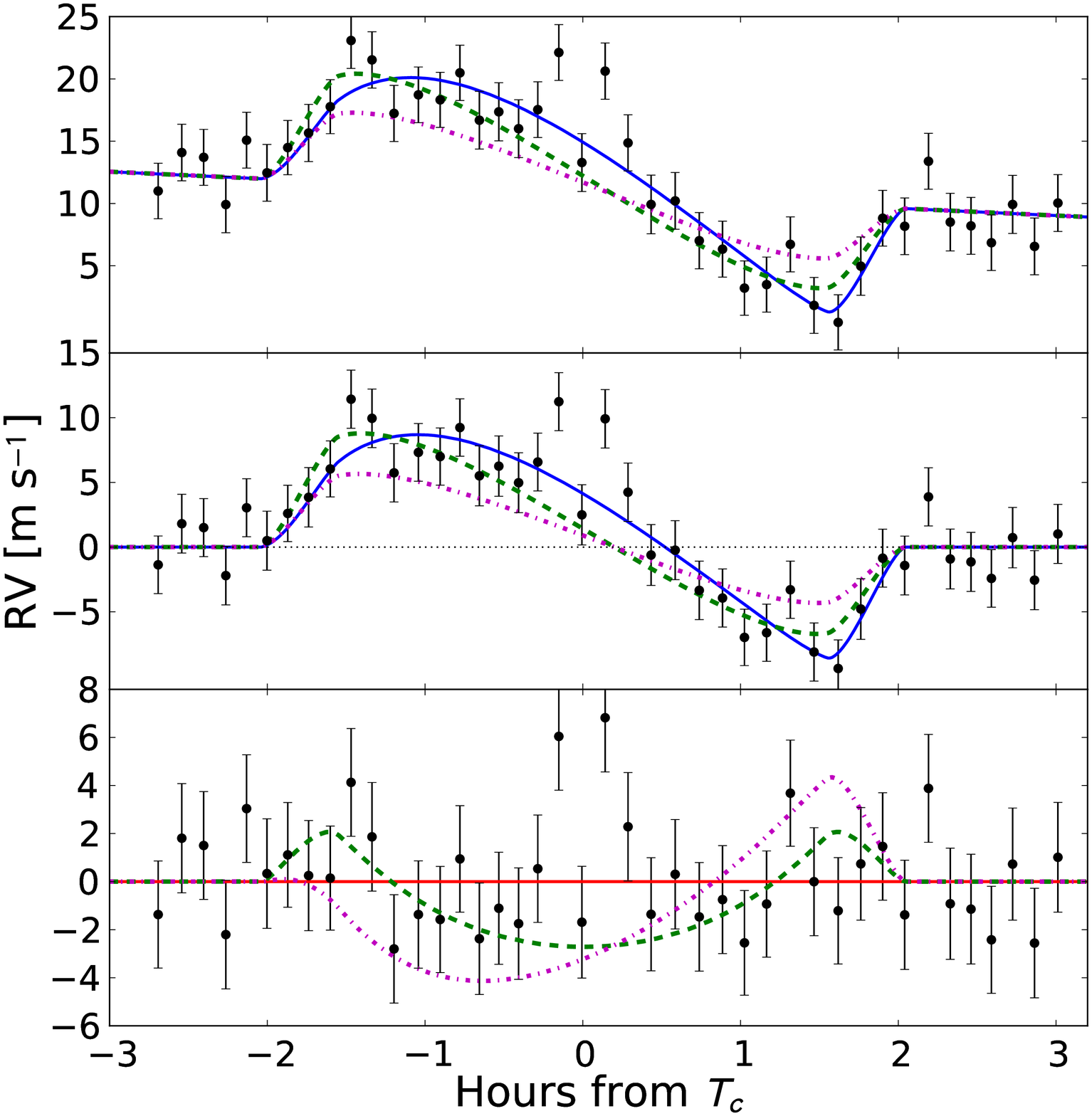}{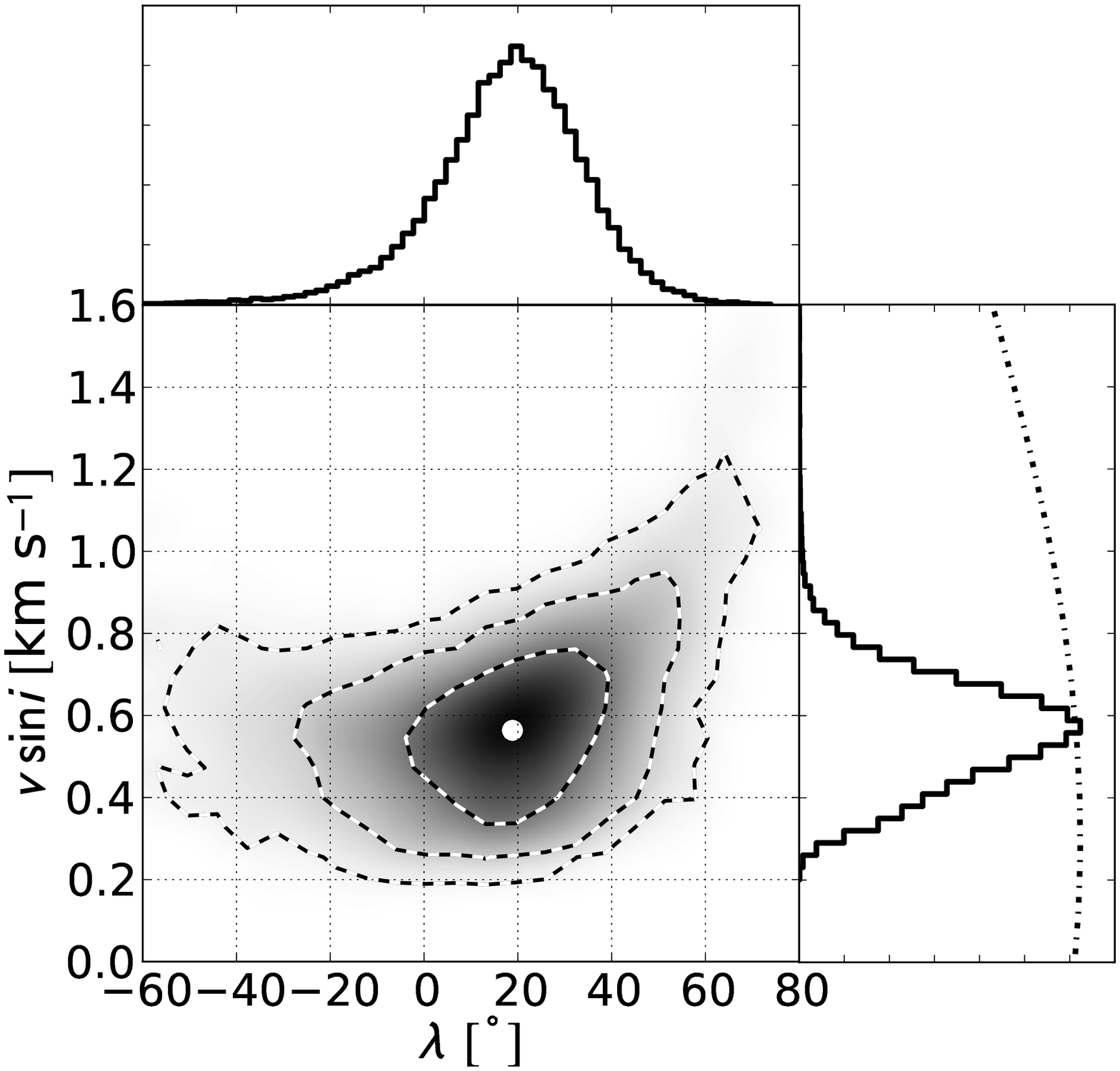}
\centering
\caption{
\emph{Left:} RV variation during the transit of HAT-P-17b due to the RM effect with best-fit model found from $\chi^{2}$ minimization overplotted. The upper panel includes the RV variation due to the orbital motion of HAT-P-17b, the middle panel shows the data and model with the orbital motion removed, and the bottom panel shows the residuals to the model. The solid blue line represents our adopted model including all line broadening effects and the convective blueshift, the dashed green line shows the model without the convective blueshift, and the dot-dashed magenta line shows an idealized model in which the line profiles are described only by rotational broadening, similar to the approach of \citet{Ohta05}. Note the asymmetry in the RM curve caused by a combination of the convective blueshift and a slight misalignment. Three data points to the right of the x-axis limit were included in the modeling, but omitted from the plot for clarity.
\emph{Right:} Posterior distribution of $\vsini$ vs. $\lambda$ from the DE-MCMC analysis of the spectroscopic transit. The dashed lines are 68\%, 95\%, and 99\% confidence intervals and the white dot is the median value of the distribution. The dot-dashed line in the right histogram shows the $0.3 \pm 1.5$ \kms prior on $v\ \sin{i}$ from the SME analysis of H12.
}
\label{rm}
\end{figure*}

\tablenum{3}
\begin{deluxetable}{lrr}[h]
\label{tab:rm}
\tablewidth{0pt} 
\tabletypesize{\small}
\tablecaption{Rossiter-McLaughlin MCMC results}
\tablehead{\colhead{Parameter} & \colhead{Value} & \colhead{Units}}
\tablehead{\colhead{Parameter} & \colhead{Value} & \colhead{Units}}
\startdata
\sidehead{\bf {RM Step Parameters:}}
log($P$) & 1.0144585 $\pm 3.8e-07$ & log(days)\\
$T_{\rm tra}$ & 2456165.8551 $\pm 0.0011$ & \bjdtdb\\
$\sqrt{e}\cos{\omega}$ & -0.544 $^{+0.007}_{-0.0068}$ & \\
$\sqrt{e}\sin{\omega}$ & -0.214 $\pm 0.015$ & \\
log($R_{p}/R_{\star})$ & -0.9073 $\pm 0.0032$ & \\
log($a/R_{\star}$) & 1.3531 $^{+0.0082}_{-0.0085}$ & \\
$\cos{i}$ & 0.0123 $^{+0.0029}_{-0.0032}$ & \\
$u_{1}+u_{2}$ & 0.736 $^{+0.097}_{-0.095}$ & \\
$\sqrt{v\sin{i}}\cos{\lambda}$ & 0.687 $^{+0.076}_{-0.1}$ & $\sqrt{\mbox{km s}^{-1}}$\\
$\sqrt{v\sin{i}}\sin{\lambda}$ & 0.24 $^{+0.19}_{-0.2}$ & $\sqrt{\mbox{km s}^{-1}}$\\
$\beta$ & 4.1 $^{+2.6}_{-2.3}$ & \kms\\
$\gamma_{H}$ & $\equiv$ 0.9  & \kms\\
$\zeta$ & $\equiv$ 4.8  & \kms\\
$v_{CB}$ & -0.65 $\pm 0.23$ & log(\kms)\\
$v_{CM}$ & 11.13 $^{+0.63}_{-0.6}$ & \ms\\
$\dot{\gamma}_{RM}$ & -17.5 $^{+4.8}_{-4.9}$ & \ms day$^{-1}$\\
log(jitter) & 0.289 $^{+0.077}_{-0.083}$ & log(\ms)\\
\\
\hline
\sidehead{\bf {RM Model Parameters:}}
$P$ & 10.3385231 $^{+9.1e-06}_{-9.2e-06}$ & days\\
$T_{\rm tra}$ & 2456165.8551 $\pm 0.0011$ & \bjdtdb\\
$e$ & 0.342 $^{+0.0046}_{-0.0047}$ & \\
$\omega$ & 201.5 $^{+1.5}_{-1.6}$ & degrees\\
$R_{p}/R_{\star}$ & 0.12378 $^{+0.00092}_{-0.00091}$ & \\
$a/R_{\star}$ & 22.55 $^{+0.43}_{-0.44}$ & \\
$i$ & 89.3 $^{+0.18}_{-0.17}$ & degrees\\
$u_{1}$ & 0.575 $^{+0.048}_{-0.047}$ & \\
$v\sin{i}$ & 0.56 $^{+0.12}_{-0.14}$ & \kms\\
$\lambda$ & 19 $^{+14}_{-16}$ & degrees\\
$\beta$ & 4.1 $^{+2.6}_{-2.3}$ & \kms\\
$\gamma_{H}$ & $\equiv$ 0.9  & \kms\\
$\zeta$ & $\equiv$ 4.8  & \kms\\
$v_{CB}$ & -0.65 $\pm 0.23$ & \kms\\
$v_{CM}$ & 11.13 $^{+0.63}_{-0.6}$ & \ms\\
$\dot{\gamma}_{RM}$ & -17.5 $^{+4.8}_{-4.9}$ & \ms day$_{-1}$\\
jitter & 1.95 $^{+0.38}_{-0.34}$ & \ms\\
\enddata
\end{deluxetable}

%%%%%%%%%%%%%%%%%%%%%%%%%%%%%%%%%%%%%%%%%%%%%%%%%%%%%
\subsection{Spectroscopic transit}
\label{RM}
At first glance the RM data follow the typical redshift then blueshift pattern of a spin-orbit-aligned system. However, the data do not cross zero until slightly after the time of mid-transit. The small asymmetry in the RM curve (Figure \ref{rm}) suggests a slight misalignment. We also used ExoPy to analyze the spectroscopic transit data. Our model of the RM effect takes the form of

\begin{equation}
\label{eqn:rm}
RM_{\mbox{net}}(t) = \Delta v(t) + V_{CB}(t) + S(t - T_{\rm tra}) + v_{CM}
\end{equation}
where $\Delta v$(t) is given by equation 16 of \citet{Hirano11} and is discussed in more detail in section \ref{hirano} below. $V_{CB}(t)$ is the anomalous radial velocity shift due to the convective blueshift \citep[][discussed in section \ref{CB} below]{Shporer11}, $\dot{\gamma}_{RM}$ is the radial velocity slope observed during transit due to the orbital motion of HAT-P-17b+c, $t$ are the flux-weighted exposure mid-times of the observations in \bjdtdb, $T_{\rm tra}$ is the \bjdtdb\ of mid-transit, and $v_{CM}$ is an arbitrary additive constant velocity. $P_{b}$ is constrained to the value obtained in the RV analysis, and $T_{\rm tra,b}$ is constrained by propagating the error on $P_{b}$ and $T_{\rm tra,b}$ found from the Keplerian analysis to the transit epoch of the night of 2012 Aug 26 ($T_{\rm tra,b} = 2456165.8553 \pm 0.0012$ \bjdtdb). The amplitude of the \hjdutc\ to \bjdtdb\ correction applied to the RV data was $\sim$67 seconds, or about half of the propagated uncertainty on the mid-transit time which highlights the importance of working in a standardized and consistent time system. The same stellar jitter that contributes to the scatter in the residuals to our Keplerian orbital fit can be seen as systematic trends on shorter timescales, and allowing $\dot{\gamma}_{RM}$ and $v_{CM}$ to be free parameters in the fit prevents these trends from biasing the obliquity measurement. We refer the reader to \citet{Albrecht12a} for a detailed discussion of the effect of stellar jitter on obliquity measurements via the RM effect.

%%%%%%%%%%%%%%%%%%%%%%%%%%%%%%%%%%%%%%%%%%%%%%%%%%%%%
\subsubsection{Semi-analytical Rossiter-McLaughlin model}
\label{hirano}
The shape and amplitude of $\Delta v$ depends on nine parameters. Five describe the decrease in flux as the planet transits its host star; the planet to star radius ratio ($R_{p}/R_{\star}$), the semi-major axis of the orbit in units of stellar radii ($a/R_{\star}$), the inclination of the orbit relative to our line-of-sight ($i$), and two quadratic limb darkening coefficients ($u_{1}$,$u_{2}$). We assigned Gaussian priors to $P_{b}$, $T_{\rm tra,b}$, $R_{p}/R_{\star}$, $a/R_{\star}$, and $i$ from the values given in H12 as these are poorly constrained by the RM data alone.

Two more geometrical parameters contribute to the shape of the spectroscopic transit; the rotational velocity of the star projected onto the plane of the sky ($v\sin{i_{\star}}$), and the angle between the rotational axis of the star projected onto the plane of the sky and the planet's orbital angular momentum vector ($\lambda$). We adopt a value of $v\sin{i_{\star}} = 0.3 \pm 1.5$ \kms\ as a Gaussian prior that was obtained from the Spectrocospy Made Easy \citep[SME,][]{Valenti96} analysis performed in H12.

Some of the orbital parameters from the RV analysis also have a small effect on the timing and duration of the transit. We assigned Gaussian priors to $e_{b}$ and $\omega_{b}$ from the results of the RV analysis.

The semi-analytical model of \citet{Hirano11} also includes three parameters that describe the sources of line broadening ($\beta$, $\gamma_{H}$, and $\zeta$). Together with the rotational broadening of the star, these parameters provide a realistic analytical description of the observed line profiles in the spectra. $\beta$ includes both the Gaussian instrumental line profile and the Gaussian dispersion from micro-tubulence. We adopted a fixed value of 3.0 \kms\ for $\beta$ that is the result of summing in quadrature the width of the HIRES PSF (2.2 \kms) and 2.0 \kms\ micro-turbulence broadening profile \citep{Albrecht12}. $\gamma_{H}$ is the Lorentzian dispersion of the spectral lines primarily due to pressure broadening. We adopted a fixed value of 0.9 \kms\ that was found to be a good match to the HIRES spectra of several stars \citep{Hirano11}. The most significant of the line profile parameters is the macroturbulence broadening ($\zeta$). We used equation \ref{eqn:macro} from \citet{Valenti05}, 
\begin{equation}
\label{eqn:macro}
\zeta = \left(3.98 - \frac{T_{\text{eff}} - 5770\ \mbox{K}}{650\ \mbox{K}}\right)\ \mbox{km\ s$^{-1}$}
\end{equation}
and \Teff~$= 5246$ K \citep{Howard12} to calculate a value of 4.8 \kms and assigned a conservative Gaussian prior of 3.0 \kms\ in the DE-MCMC analysis. We found that changing the prior centers on $\beta$, $\gamma_{H}$, and $\zeta$ had little effect on the resulting posterior distributions of $\lambda$, and for this reason we also could not remove the Gaussian prior on $\zeta$ without the DE-MCMC chains wandering into unphysical regions of parameter space.

%%%%%%%%%%%%%%%%%%%%%%%%%%%%%%%%%%%%%%%%%%%%%%%%%%%%%
\subsubsection{Convective blueshift}
\label{CB}
The convective blueshift ($V_{CB}(t)$) is caused by the net convective motion of the stellar photosphere. Hotter material from below the photosphere rises upward towards the observer due to convection and is only partially canceled by downwelling cold material, causing a net blueshift of order 1 \kms. Since we are only interested in relative radial velocities this net blueshift is unimportant. However, because the convective blueshift is strongest near the center of the star and weaker near the limbs, the transiting planet occults areas of the star that have different contributions to the net convective blueshift. This causes a time-varying component of the convective blueshift during the spectroscopic transit of order 2 \ms. We refer the reader to \citet{Shporer11} for a more detailed discussion of the convective blueshift, and its influence on the measurement of $\lambda$. Since the $\vsini$ of HAT-P-17 is relatively low, the amplitude of the spectroscopic transit signal is only about 7 \ms and thus the convective blueshift is a significant effect and must be included in our model. We found that adding the convective blueshift changes the measurement of $\lambda$ by $\sim$1 $\sigma$, pushing it towards zero when the $V_{CB}(t)$ is included.

We used a numerical model based on the work by \citet{Shporer11} similar to the approach used by \citep{Albrecht12a}. We made an initial assumption that the convective blueshift is similar to that of the sun to create a model grid for a range of $R_{p}/R_{\star}$ and impact parameters. We interpolated this grid at each step in the DE-MCMC chains. We left the velocity of the photosphere ($v_{CB}$) as a free parameter
%with a conservative Gaussian prior of $-0.5 \pm 1.5$ \kms\
to account for the differences between HAT-P-17 and the sun. By definition we expect $v_{CB}$ to be negative and for this reason we rejected models with positive $v_{CB}$ in the DE-MCMC chains. Note the difference between the time-dependent RV signal caused by the convective blueshift ($V_{CB}(t)$) and the fitted scaling factor ($v_{CB}$).

%- Anything else? (Simon)

\subsubsection{Results}
The results of the RM modeling are presented in Table \ref{tab:rm}. Figure \ref{rm} shows the spectroscopic transit data with the best-fitting model overplotted and the resulting posterior distributions of $v\sin{i_{\star}}$ and $\lambda$. We measure the sky-projected angle between the orbital angular momentum vector and the stellar rotation axis to be $\lambda = 19^{+14}_{-16}$ degrees. This indicates that planet b's orbit is misaligned with the stellar rotation at a confidence level of only 1.2 $\sigma$. Our value of $\vsini=0.54 \pm 0.15$ \kms\ is slightly larger than the value reported in H12 ($\vsini=0.3\pm0.5$ \kms), but well within the 1 $\sigma$ uncertainty from the SME analysis.

We experimented with fixing $\vsini$, $v_{CB}$, and the transit parameters and saw no significant changes in the resulting posterior distribution of $\lambda$. When we neglect the convective blueshift in our model we measure a much more significant (presumably artificial) misalignment with $\lambda=37\pm12$ degrees. We also examined the diagnostics from the Doppler analysis of the two outliers on either side of the mid-transit. We found no evidence of systematic errors, poor fits, or other reasons to doubt the integrity of these model outlier points. Removing them from our fit did not change the results other than decreasing the reduced $\chi{^2}$.

\subsection{Additional test for misalignment}
%We also used the method of \citet{Schlaufman10} to independently look for evidence of misalignment.
We also used the method of \citet{Schlaufman10} to check for consistency with our RM modeling.This approach compares the measured $\vsini$ to an empirical estimate of the expected value of $v = 2\pi R_{\star}/P_{\rm rot}$, where $P_{\rm rot}$ is the rotation period of the star based on the mass-age-rotation relations established from observations of the Hyades and Prasepe clusters summarized by \citet{Irwin09}. If the sky-projected inclination of the stellar rotation ($i_{\star}$) is close to $90^{\circ}$ we would expect the measured $\vsini$ to closely correspond to $v$.  Since we know that the orbit of planet b is viewed nearly edge on ($i = 89.3^{+0.18}_{-0.17}$ degrees), an observed $\vsini$ significantly different than  $v$ would suggest spin-orbit misalignment. We use

\begin{equation}
P_{\star}(M_{\star},\tau_{\star}) = P_{\star,0}(M_{\star})\left(\frac{\tau_{\star}}{650 \text{ Myr}}\right)^{1/2}
\end{equation}
from \citet{Schlaufman10} to calculate the expected rotation period of HAT-P-17 at the age given by H12. In the above equation, $P_{\star,0}(M_{\star})$ is the rotation period of the star as a function of mass at an age of 650 Myr (12 days), and $\tau_{\star}=7.8\pm3.3$ Gyr is the current age of the star. Assuming that the uncertainty in the age of HAT-P-17 is the dominant source of uncertainty, we calculate $P_{\star}(0.857 M_{\odot}, 7.8\pm3.3 \text{Gyr}) = 42 ^{+8}_{-10}$ days. For $i_{\star} = 90^{\circ}$ we calculate $v = 1.0^{+0.4}_{-0.2}$ \kms.
%Our measured value of $\vsini=0.56^{+0.12}_{-0.13}$ is significantly ($\sim$2$~\sigma$) lower than $v$, suggesting that the star is inclined relative to our line of sight ($\sin{i_{\star}} < 1$).
%Since we know the orbital inclination of planet b relative to our line of sight is very close to $90^{\circ}$ ($i=89.29^{+0.18}_{-0.17}$ degrees) the Schlaufman test gives independent evidence for spin-orbit misalignment.

We compared our measured $\vsini=0.56^{+0.12}_{-0.14}$ from the RM analysis to $v$ by calculating the rotation statistic ($\Theta$) from \citet{Schlaufman10} defined as:

\begin{equation}
\Theta \equiv \frac{v - \vsini_{\rm obs}}{\sqrt{\sigma^{2}_{v} + \sigma^{2}_{\rm obs}}}
\end{equation}
where 
%$v$ is the calculated assuming $i_{\star}=90^{\circ}$, 
$\vsini_{\rm obs}$ is measured from the RM analysis, and $\sigma_{v}$ and $\sigma_{\rm obs}$ are the uncertainties on $v$ and $\vsini$ respectively.
The difference between $\vsini$ and $v$ may suggest that the stellar rotation is inclined with respect to our line of sight. However, the value of $\Theta$ = 1.9 (equivalent to 1.9 $\sigma$) is below the threshold for misalignment as defined by Schlaufman ($\Theta > 2.9$). This threshold accounts for the scatter in the empirical mass-age-rotation calibration which makes our determination of the initial rotation period of HAT-P-17 uncertain. 
The Schlaufman method provides weak, independent evidence of spin-orbit misalignment that is consistent with our obliquity measurement from the RM analysis.  However, these two low significance measurements (each less than 2 $\sigma$) do not conclusively show that the system is misaligned.
%The Schlaufman method provides a second independent source of evidence suggesting a slight misalignment, but its significance is similar to that of the RM analysis.
%distinguish between an aligned state ($i_{\star} = 0^{\circ}$) and a slightly misaligned state ($i_{\star} = 19^{\circ}$).  The value for $\Theta$ suggests that HAT-P-17 is not viewed pole on ($i_{\star} = 90^{\circ}$).
%We find that $\Theta = 1.4$ which indicates that a large misalignment is unlikely.

\subsection{Adaptive optics imaging}
\label{sec:AO}

\begin{figure}[h]
\epsscale{2.2}
\plottwo{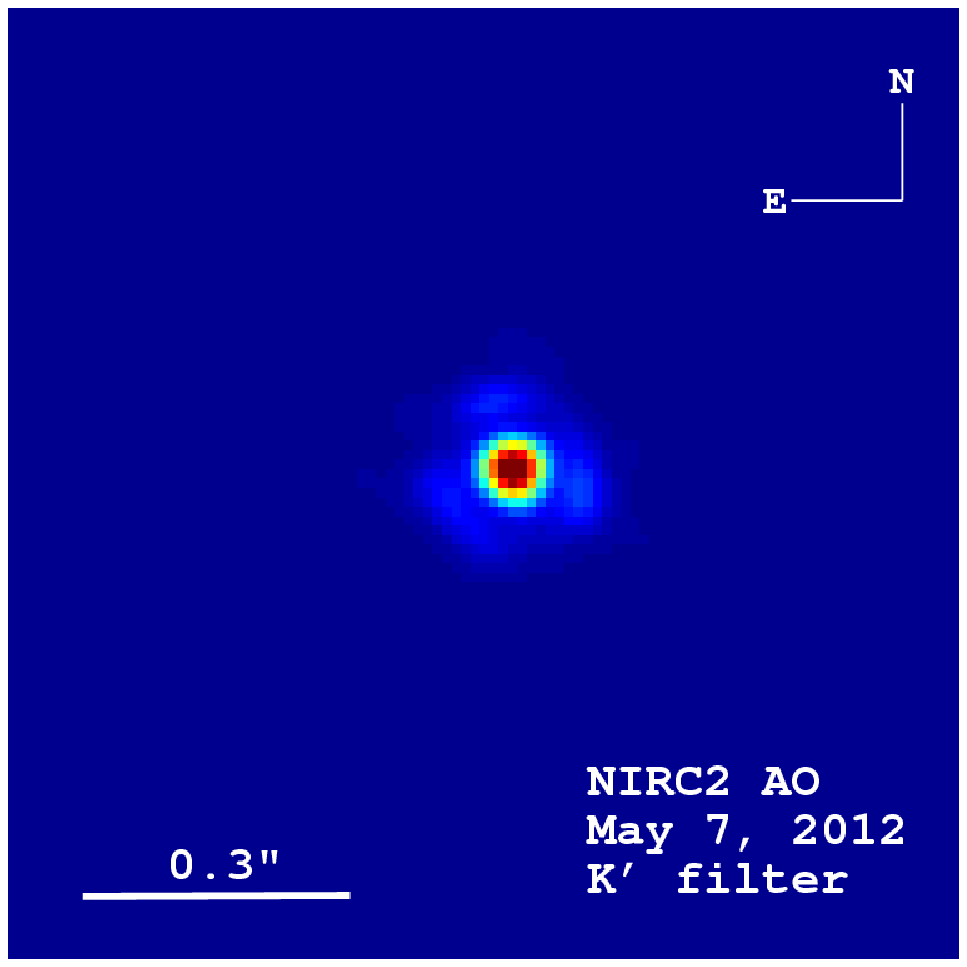}{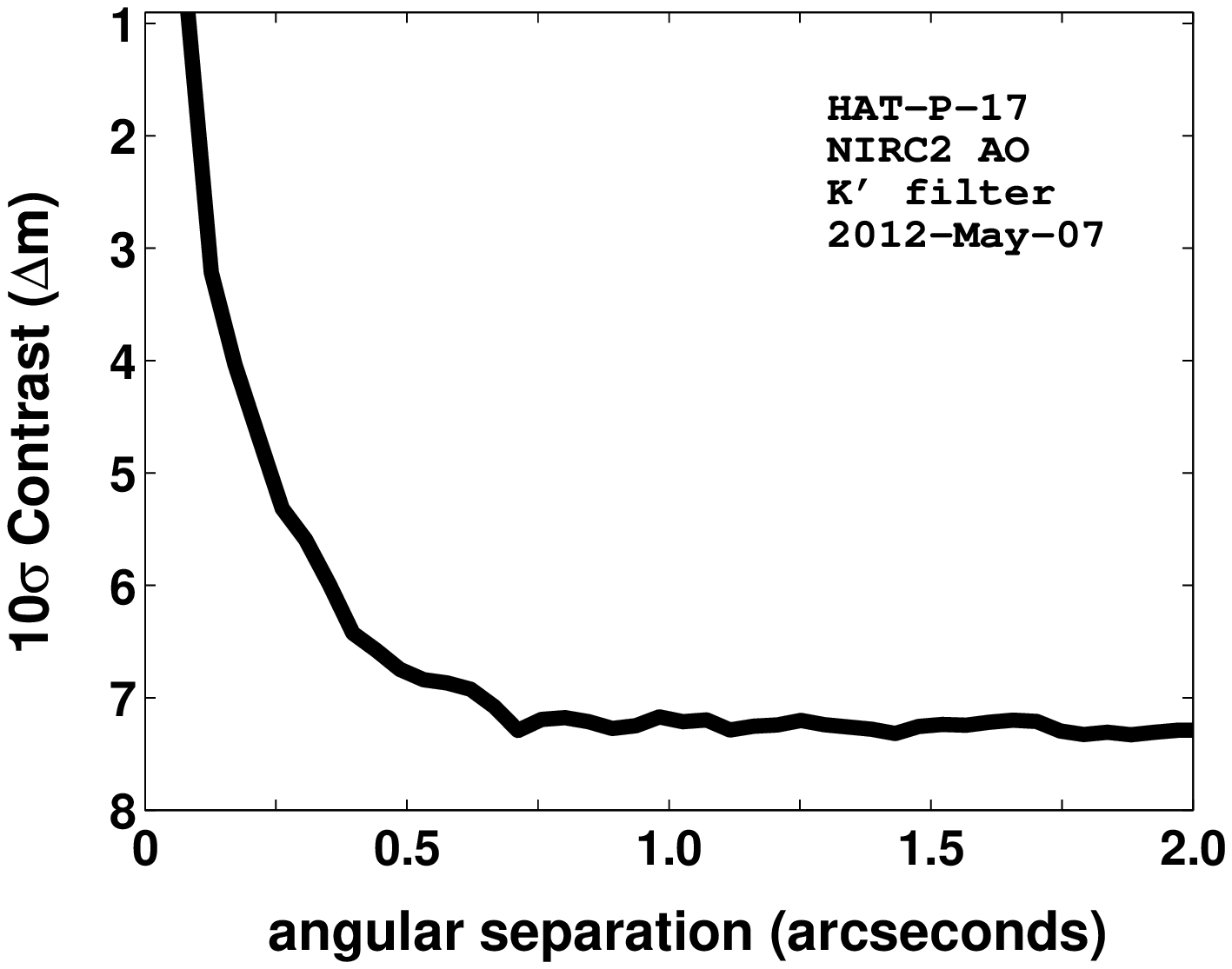}
\centering
\caption{
\emph{Top:} Keck/NIRC2 adaptive optics image.
%Contrast curve from the Keck/NIRC2 data. The black curve traces the maximum delta magnitude relative to HAT-P-17 for which a point source companion could be detected ($10\sigma$) as a function of angular separation. Companions with parameters that fall below the line would not have been detected.
\emph{Bottom:}
contrast achieved based on the final reduced AO image. Our diffraction-limited observations rule out the presence of companions with $\Delta m > 7$ mags for separations beyond $\approx0.7\arcsec$. 
}
\label{contrast}
\end{figure}

\begin{figure}[h]
\epsscale{2.5}
\plottwo{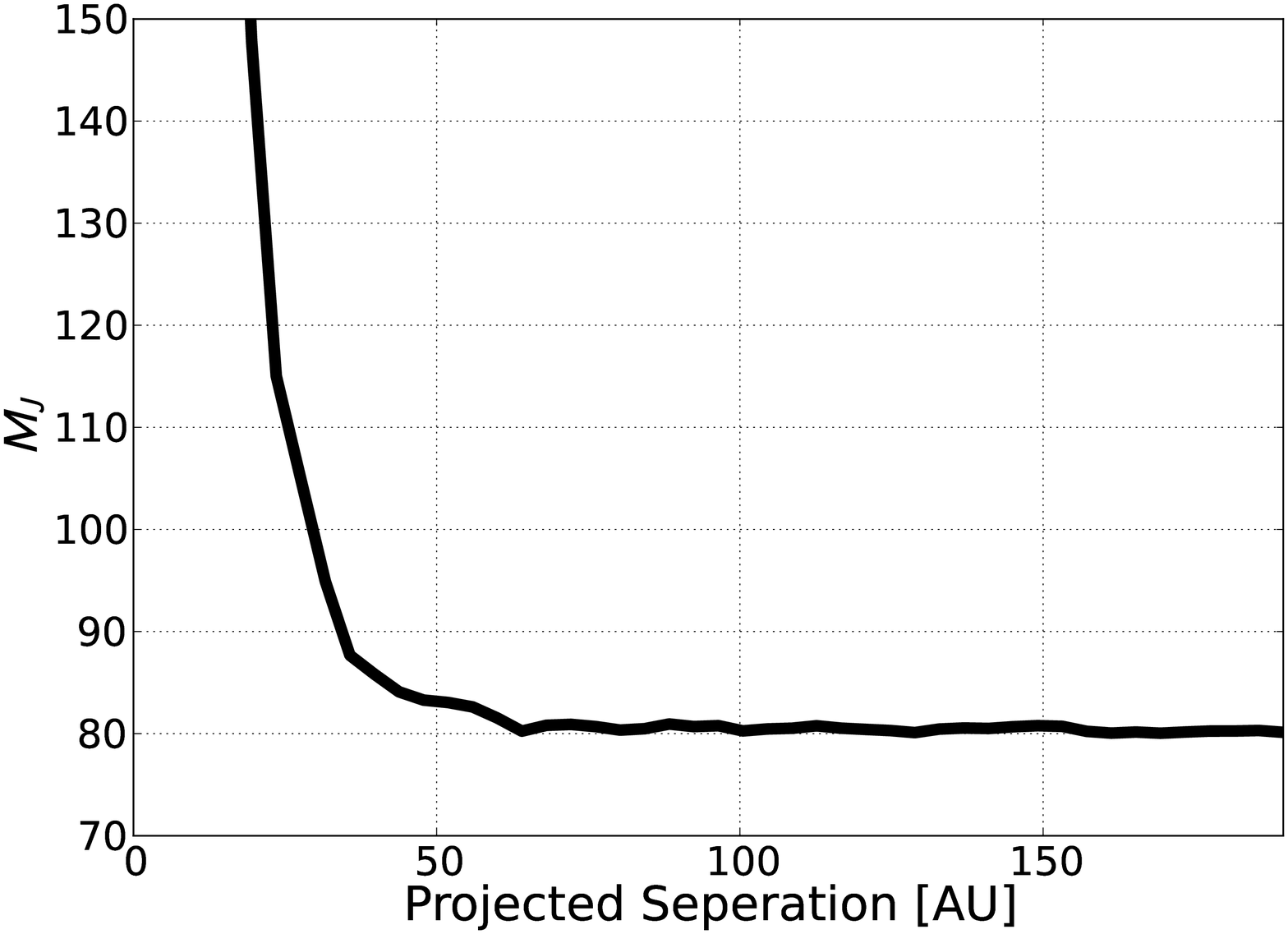}{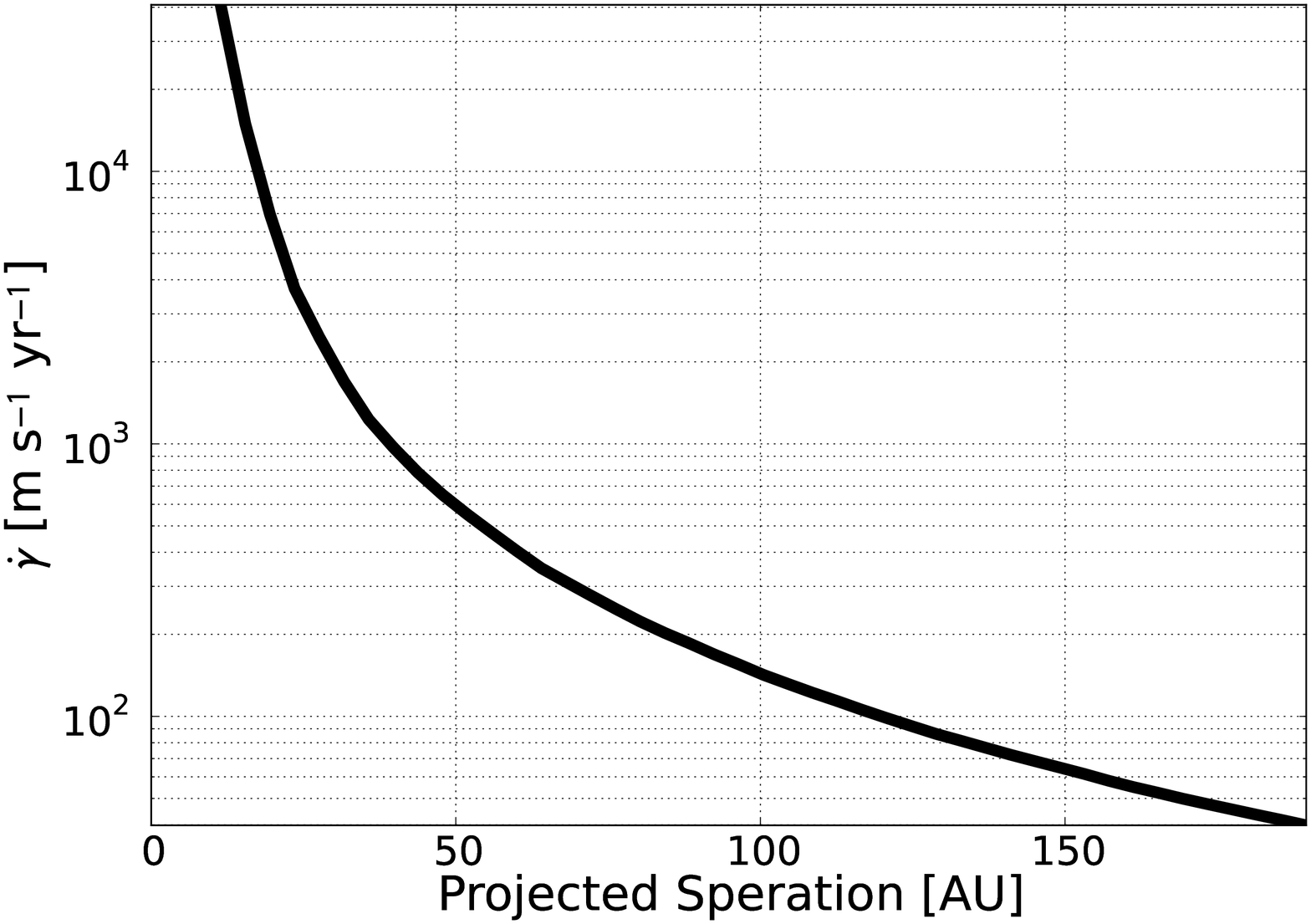}
\centering
\caption{
\emph{Top:}
We convert the measured contrast to a mass-sensitivity curve using the Baraffe et al. 2003 evolutionary models. With an age of 7.8 Gyr, we could have detected any stellar companions ($M>80M_J$) associated with HAT-P-17 at projected separations beyond $\approx 60$ AU.
%Contrast curve from figure \ref{contrast} converted into minimum mass vs. projected orbital distance using the stellar models of \citet{Baraffe02} and the distance to HAT-P-17 (90 pc, H12). A companion in the region below the line would be undetected in the AO image.
\emph{Bottom:} Same as \emph{top} converted into a predicted RV slope using Equation \ref{eqn:gammadot}. The region below the line is allowed by the data. The 3 $\sigma$ slope constraint ($|\dot{\gamma}| \leq$ 19 \ms yr$^{-1}$) from the RV analysis is just below the lower y-axis limit.
}
\label{gammadot}
\end{figure}

We carried out high resolution and high contrast imaging with adaptive optics to check for near-by companions in the context of understanding the architecture of the HAT-P-17 planetary system. Such companions are important in understanding the orbital evolution of the system. We find no evidence for off-axis sources in the immediate vicinity of HAT-P-17. To estimate our sensitivity to faint companions, we calculated the average contrast level achieved as a function of angular separation. Specifically, we compared the peak stellar intensity to the standard deviation ($\sigma$) in scattered light within a square box of width 3 FWHM, where FWHM is the PSF full-width at half-maximum (also the size of a speckle). The standard deviation is evaluated at numerous locations and the results are azimuthally averaged to create a contrast radial profile. 

We converted the contrast curve into a minimum detectable mass as a function of projected orbital separation (figure \ref{gammadot}, left panel) by interpolating the models of \citet{Baraffe02} at the age and distance of HAT-P-17 from the analysis of H12. Assuming a circular orbit and $M_{P} \ll M_{\star}$, an order-of-magnitude approximation for the maximum RV slope caused by a fourth body in the system is given by \citep{Winn09}: 
\begin{equation}
\label{eqn:gammadot}
\dot{\gamma} \approx \frac{GM_{c}\sin{i_c}}{a_{c}^{2}}.
\end{equation}
We used this approximation to find the RV slope that would be produced by a planet at the minimum detectable mass as a function of projected orbital separation (figure \ref{gammadot}, right panel). The minimum detectable mass at large separations is $\sim80M_{J}$ (coincident with the hydrogen-burning limit), far larger than the range of masses that are allowed by our RV fit and does not provide a good constraint on the orbit of planet c or a fourth companion. However, the RV data could still allow for a long-period companion in a nearly face-on orbit or one that is currently near apsis which would minimize the radial velocity slope. The AO data help us rule out these scenarios for stellar/brown dwarf companions outside $\sim50$ AU.

%%%%%%%%%%%%%%%%%%%%%%%%%%%%%%%%%%%%%%%%%%%%%%%%%%%%%
\section{Discussion}
\label{discussion}
%%%%%%%%%%%%%%%%%%%%%%%%%%%%%%%%%%%%%%%%%%%%%%%%%%%%%
HAT-P-17 is a rare planetary system with a transiting hot Jupiter and a long-period companion (HAT-P-17c). We have shown that the orbit of planet c is poorly constrained with the current RV data. We will not be able to conclusively measure the orbital parameters until a significant portion of the orbit has been observed. We find no evidence to suggest the presence of a massive 4th body. We modeled the RM effect of planet b and measure a possible misalignment of the projected plane of the orbit and the rotation axis of the host star.

\begin{figure}[h]
\epsscale{1.2}
%\plottwo{figures/tau1.pdf}{figures/tau2.pdf}
\plotone{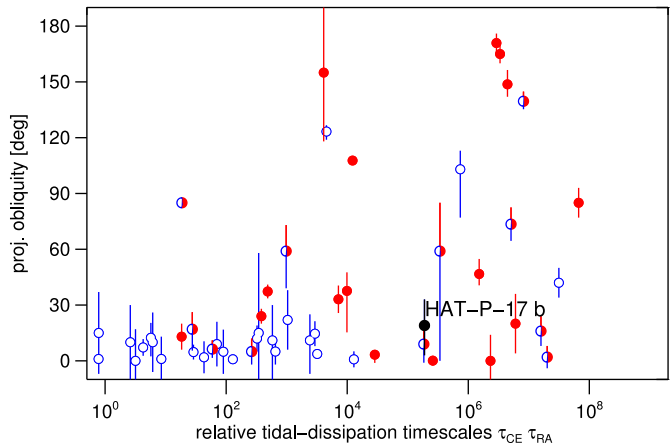}
\centering
\caption{
\emph{Top:} measured projected obliquity as a function of the alignment timescale calibrated from binary studies \citep{Albrecht12}. Stars with temperatures higher then 6250 K are shown with red filled symbols. Blue open symbols show stars with temperatures lower then 6250 K. Stars for which measured effective temperatures include 6250 K in their 1-$\sigma$ interval are shown by split symbols. We are computing the relative tidal dissipation rates as a function of stellar type, planet-to-star mass ratio, and orbital distance, using the scaling relations presented by Albrecht et al. (2012). Note that both timescales have been divided by $5\times10^{9}$.
%\emph{Bottom:} Measured projected obliquity as a function of the alignment timescale estimated from the mass of the convective envelope (eqn. 4 of \citet{Albrecht12} divided by age.
%THESE PLOTS NEED TO BE UPDATED TO MATCH THE PARAMETERS IN TABLE \ref{tab:rm}
}
\label{fig:timescale}
\end{figure}

Our constraints on a long-term RV trend (in addition to the two planet model) give an upper limit to the mass of a 4th companion of $M_{d}\sin{i_{d}}\left(\frac{a_{d}}{10 \rm AU}\right)^{2} < 10 M_{J}$ with the assumptions that the potential 4th companion is currently near a time of conjunction in a circular orbit.  The lack of companions seen in the adaptive optics image provides complimentary evidence against the presence of a 4th body more massive than $\sim$80M$_{J}$ at separations larger than $\sim$50 AU for a wide range of orbital configurations.
%The AO data suggest that a 4th body more massive than $\sim$80 M$_{J}$ at separations larger than $\sim$50 AU is unlikely, and the absence of a trend in the RV data suggests that there are no additional Jovian-mass planets in nearly edge-on circular orbits.

Given that the period of planet c reported by H12 was underestimated we do not want to over-interpret any of planet c's parameters. Instead we urge the community to continue observing this interesting system in the coming years. We will be able to assess our measured 16.8 year orbital period in $\sim$5 years when the RVs start to decrease rapidly as planet c approaches periapsis, and we will be able to start ruling out the short end of our estimate ($P_{c} \sim$ 10 years) in $\sim$3 years. If it can be shown that the system is coplanar (this would require a spectacular observational effort by searching for transits of planet c) and apsidially locked then HAT-P-17 will be of further interest because it will give us a rare opportunity to probe the interior structure of an exoplanet by measuring the tidal Love number and quality factor through dynamical modeling \citep{Batygin09,Mardling10}.

We measure the sky-projected angle between the stellar spin axis and orbital angular momentum of the inner planet (stellar obliquity) by modeling the RM effect in Keck/HIRES RV data. The RM analysis suggests a slight spin-orbit misalignment of planet b with $\sim$1.2 $\sigma$ confidence ($\lambda = 19^{+14}_{-16}$ degrees). The Schlaufman method provides additional evidence for spin-orbit misalignment, but due to the dependence on somewhat uncertain stellar evolution models and the unknown initial angular momentum of HAT-P-17 we believe that the result from our detailed RM modeling to be more robust. However, the two low-confidence measurements do not allow us to distinguish between a well-aligned system or one with a small, but non-zero, spin-orbit misalignment.

%Add two sentence paragraph restating RM and Schlaufman result + state uncertainty.  Emphasize that the RM involves detailed modeling and is therefore probably more reliable.

\citet{Winn10b} noted an emerging trend in the orbital obliquities of hot Jupiter hosts cooler than 6250 K being generally well-aligned, and hot Jupiters around stars hotter than 6250 K having a wide distribution of $\lambda$. HAT-P-17 is a cool star around which we would normally expect to find spin-orbit-aligned hot Jupiters, but due to the relatively wide orbit of HAT-P-17b we do not expect that the planet would have had time to align itself with the spin of the host star \citep{Winn10b,Albrecht12} if it were perturbed into a misaligned orbit. This is consistent with our findings of a marginally non-zero $\lambda$, however a more significant spin-orbit misalignment would provide stronger evidence against the disk-migration formation scenario. Figure \ref{fig:timescale} compares the timescales for alignment of this system with the systems studied in \citet{Albrecht12}. HAT-P-17b lies in a region of Figure \ref{fig:timescale} that shows large scatter in $\lambda$ due to the longer realignment timescales. We expect that the planets in this region of the plot retain their spin-orbit angle from the time shortly after their migration because the tidal interactions are too small to force a realignment over the age of the star. If misaligned, this is the first multi-planet system in which a spin-orbit misalignment has been measured. We also note that our measurement of $\lambda$ is only slightly less likely to be consistent with zero, and coplanarity of planet c's orbit would provide strong evidence that the system migrated quiescently in the protoplanetary disk.

For a low-amplitude RM system like HAT-P-17, we find that the convective blueshift is an important effect that must be included in our model for an accurate measurement of $\lambda$.
%Models that neglect the convective blueshift overestimate $\lambda$ relative to HAT-P-17b.
With 2-3 additional measurements of the RM effect we should be able to conclusively ($\sim$3 $\sigma$) determine if the system is misaligned which will help us understand the formation of the HAT-P-17 system and other similar systems.

%%%%%%%%%%%%%%%%%%%%%%%%%%%%%%%%%%%%%%%%%%%%%%%%%%%%%
%\section{Summary}
%\label{summary}
%%%%%%%%%%%%%%%%%%%%%%%%%%%%%%%%%%%%%%%%%%%%%%%%%%%%%

%%%%%%%%%%%%%%%%%%%%%%%%%%%%%%%%%%%%%%%%%%%%%%%%%%%%%
\acknowledgments{
We thank the referee for his/her prompt response, careful reading, and useful comments, Jason Eastman for providing the EXOFAST code to community, and the many observers who contributed to the measurements reported here.  We gratefully acknowledge the efforts and dedication of the Keck Observatory staff, especially Scott Dahm, Greg Doppman, Hien Tran, and Grant Hill for support of HIRES and Greg Wirth for support of remote observing.  Finally, we extend special thanks to those of Hawai`ian ancestry on whose sacred mountain of Mauna Kea we are privileged to be guests.  Without their generous hospitality, the Keck observations presented herein would not have been possible.
}

%%%%%%%%%%%%%%%%%%%%%%%%%%%%%%%%%%%%%%%%%%%%%%%%%%%%%
%% Facilities
{\it Facilities:}
\facility{KeckI:HIRES}
\facility{KeckII:NIRC2}
%%%%%%%%%%%%%%%%%%%%%%%%%%%%%%%%%%%%%%%%%%%%%%%%%%%%%

%%%%%%%%%%%%%%%%%%%%%%%%%%%%%%%%%%%%%%%%%%%%%%%%%%%%%
% REFERENCES
%\clearpage
\bibliographystyle{apj}
\bibliography{references}

\begin{thebibliography}{38}
\expandafter\ifx\csname natexlab\endcsname\relax\def\natexlab#1{#1}\fi

\bibitem[{{Albrecht} {et~al.}(2012a){Albrecht}, {Winn}, {Butler}, {Crane},
  {Shectman}, {Thompson}, {Hirano}, \& {Wittenmyer}}]{Albrecht12a}
{Albrecht}, S., {Winn}, J.~N., {Butler}, R.~P., {Crane}, J.~D., {Shectman},
  S.~A., {Thompson}, I.~B., {Hirano}, T., \& {Wittenmyer}, R.~A. 2012a, \apj,
  744, 189

\bibitem[{Albrecht {et~al.}(2012b)Albrecht, Winn, Johnson, Howard, Marcy,
  Butler, Arriagada, Crane, Shectman, Thompson, Hirano, Bakos, \&
  Hartman}]{Albrecht12}
Albrecht, S., {et~al.} 2012b, \apj, 757, 18

\bibitem[{{Bakos} {et~al.}(2004){Bakos}, {Noyes}, {Kov{\'a}cs}, {Stanek},
  {Sasselov}, \& {Domsa}}]{Bakos04}
{Bakos}, G., {Noyes}, R.~W., {Kov{\'a}cs}, G., {Stanek}, K.~Z., {Sasselov},
  D.~D., \& {Domsa}, I. 2004, \pasp, 116, 266

\bibitem[{{Bakos} {et~al.}(2009){Bakos}, {Howard}, {Noyes}, {Hartman},
  {Torres}, {Kov{\'a}cs}, {Fischer}, {Latham}, {Johnson}, {Marcy}, {Sasselov},
  {Stefanik}, {Sip{\H o}cz}, {Kov{\'a}cs}, {Esquerdo}, {P{\'a}l},
  {L{\'a}z{\'a}r}, {Papp}, \& {S{\'a}ri}}]{Bakos09}
{Bakos}, G.~{\'A}., {et~al.} 2009, \apj, 707, 446

\bibitem[{{Baraffe} {et~al.}(2002){Baraffe}, {Chabrier}, {Allard}, \&
  {Hauschildt}}]{Baraffe02}
{Baraffe}, I., {Chabrier}, G., {Allard}, F., \& {Hauschildt}, P.~H. 2002, \aap,
  382, 563

\bibitem[{{Batygin} {et~al.}(2009){Batygin}, {Bodenheimer}, \&
  {Laughlin}}]{Batygin09}
{Batygin}, K., {Bodenheimer}, P., \& {Laughlin}, G. 2009, \apjl, 704, L49

\bibitem[{{Butler} {et~al.}(1996){Butler}, {Marcy}, {Williams}, {McCarthy},
  {Dosanjh}, \& {Vogt}}]{Butler96}
{Butler}, R.~P., {Marcy}, G.~W., {Williams}, E., {McCarthy}, C., {Dosanjh}, P.,
  \& {Vogt}, S.~S. 1996, \pasp, 108, 500

\bibitem[{{Eastman} {et~al.}(2013){Eastman}, {Gaudi}, \& {Agol}}]{Eastman12}
{Eastman}, J., {Gaudi}, B.~S., \& {Agol}, E. 2013, \pasp, 125, 83

\bibitem[{{Eastman} {et~al.}(2010){Eastman}, {Siverd}, \& {Gaudi}}]{Eastman10}
{Eastman}, J., {Siverd}, R., \& {Gaudi}, B.~S. 2010, \pasp, 122, 935

\bibitem[{{Fabrycky} \& {Tremaine}(2007)}]{Fabrycky07}
{Fabrycky}, D., \& {Tremaine}, S. 2007, \apj, 669, 1298

\bibitem[{Fabrycky {et~al.}(2012)Fabrycky, Ford, Steffen, Rowe, Carter,
  Moorhead, Batalha, Borucki, Bryson, Buchhave, Christiansen, Ciardi, Cochran,
  Endl, Fanelli, Fischer, Fressin, Geary, Haas, Hall, Holman, Jenkins, Koch,
  Latham, Li, Lissauer, Lucas, Marcy, Mazeh, McCauliff, Quinn, Ragozzine,
  Sasselov, \& Shporer}]{Fabrycky12}
Fabrycky, D.~C., {et~al.} 2012, \apj, 750, 114

\bibitem[{{Ford}(2006)}]{Ford06}
{Ford}, E.~B. 2006, \apj, 642, 505

\bibitem[{{Gelman} {et~al.}(2003){Gelman}, {Carlin}, {Stern}, \&
  {Rubin}}]{Gelman03}
{Gelman}, A., {Carlin}, J.~B., {Stern}, H.~S., \& {Rubin}, D.~B. 2003,
  {Bayesian Data Analysis}, 2nd edn. ({Chapman and Hall})

\bibitem[{Hirano {et~al.}(2011)Hirano, Suto, Winn, \& Taruya}]{Hirano11}
Hirano, T., Suto, Y., Winn, J.~N., \& Taruya, A. 2011, \apj

\bibitem[{Hirano {et~al.}(2012)Hirano, Narita, Sato, Takahashi, Masuda, Takeda,
  Aoki, Tamura, \& Suto}]{Hirano12}
Hirano, T., {et~al.} 2012, \apjl, 759, L36

\bibitem[{{Holman} {et~al.}(2006){Holman}, {Winn}, {Latham}, {O'Donovan},
  {Charbonneau}, {Bakos}, {Esquerdo}, {Hergenrother}, {Everett}, \&
  {P{\'a}l}}]{Holman06}
{Holman}, M.~J., {et~al.} 2006, \apj, 652, 1715

\bibitem[{{Holman} {et~al.}(2010){Holman}, {Fabrycky}, {Ragozzine}, {Ford},
  {Steffen}, {Welsh}, {Lissauer}, {Latham}, {Marcy}, {Walkowicz}, {Batalha},
  {Jenkins}, {Rowe}, {Cochran}, {Fressin}, {Torres}, {Buchhave}, {Sasselov},
  {Borucki}, {Koch}, {Basri}, {Brown}, {Caldwell}, {Charbonneau}, {Dunham},
  {Gautier}, {Geary}, {Gilliland}, {Haas}, {Howell}, {Ciardi}, {Endl},
  {Fischer}, {F{\"u}r{\'e}sz}, {Hartman}, {Isaacson}, {Johnson}, {MacQueen},
  {Moorhead}, {Morehead}, \& {Orosz}}]{Holman10}
---. 2010, Science, 330, 51

\bibitem[{Howard {et~al.}(2012)Howard, Bakos, Hartman, Torres, Shporer, Mazeh,
  Kov{\'a}cs, Latham, Noyes, Fischer, Johnson, Marcy, Esquerdo, B{\'e}ky,
  Butler, Sasselov, Stefanik, Perumpilly, L{\'a}z{\'a}r, Papp, \&
  S{\'a}ri}]{Howard12}
Howard, A.~W., {et~al.} 2012, \apj, 749, 134

\bibitem[{{Irwin} \& {Bouvier}(2009)}]{Irwin09}
{Irwin}, J., \& {Bouvier}, J. 2009, in IAU Symposium, Vol. 258, IAU Symposium,
  ed. {E.~E.~Mamajek, D.~R.~Soderblom, \& R.~F.~G.~Wyse}, 363--374

\bibitem[{Kipping {et~al.}(2011)Kipping, Hartman, Bakos, Torres, Latham,
  Bayliss, Kiss, Sato, B{\'e}ky, Kov{\'a}cs, Quinn, Buchhave, Andersen, Marcy,
  Howard, Fischer, Johnson, Noyes, Sasselov, Stefanik, L{\'a}z{\'a}r, Papp,
  S{\'a}ri, \& F{\H u}r{\'e}sz}]{Kipping11}
Kipping, D.~M., {et~al.} 2011, \apj, 142, 95

\bibitem[{{Liddle}(2007)}]{Liddle07}
{Liddle}, A.~R. 2007, \mnras, 377, L74

\bibitem[{{Lin} {et~al.}(1996){Lin}, {Bodenheimer}, \& {Richardson}}]{Lin96}
{Lin}, D.~N.~C., {Bodenheimer}, P., \& {Richardson}, D.~C. 1996, \nat, 380, 606

\bibitem[{{Mardling}(2010)}]{Mardling10}
{Mardling}, R.~A. 2010, \mnras, 407, 1048

\bibitem[{{Nagasawa} {et~al.}(2008){Nagasawa}, {Ida}, \& {Bessho}}]{Nagasawa08}
{Nagasawa}, M., {Ida}, S., \& {Bessho}, T. 2008, \apj, 678, 498

\bibitem[{{Naoz} {et~al.}(2011){Naoz}, {Farr}, {Lithwick}, {Rasio}, \&
  {Teyssandier}}]{Naoz11}
{Naoz}, S., {Farr}, W.~M., {Lithwick}, Y., {Rasio}, F.~A., \& {Teyssandier}, J.
  2011, \nat, 473, 187

\bibitem[{{Ohta} {et~al.}(2005){Ohta}, {Taruya}, \& {Suto}}]{Ohta05}
{Ohta}, Y., {Taruya}, A., \& {Suto}, Y. 2005, \apj, 622, 1118

\bibitem[{Schlaufman(2010)}]{Schlaufman10}
Schlaufman, K.~C. 2010, \apj, 719, 602

\bibitem[{Shporer \& Brown(2011)}]{Shporer11}
Shporer, A., \& Brown, T. 2011, \apj, 733, 30

\bibitem[{{Steffen} {et~al.}(2012){Steffen}, {Ragozzine}, {Fabrycky}, {Carter},
  {Ford}, {Holman}, {Rowe}, {Welsh}, {Borucki}, {Boss}, {Ciardi}, \&
  {Quinn}}]{Steffen12}
{Steffen}, J.~H., {et~al.} 2012, Proceedings of the National Academy of
  Science, 109, 7982

\bibitem[{Ter~Braak(2006)}]{Braak06}
Ter~Braak, C. 2006, Statistics and Computing

\bibitem[{Valenti \& Fischer(2005)}]{Valenti05}
Valenti, J.~A., \& Fischer, D.~A. 2005, \apjs, 159, 141

\bibitem[{{Valenti} \& {Piskunov}(1996)}]{Valenti96}
{Valenti}, J.~A., \& {Piskunov}, N. 1996, \aaps, 118, 595

\bibitem[{{Vogt} {et~al.}(1994){Vogt}, {Allen}, {Bigelow}, {Bresee}, {Brown},
  {Cantrall}, {Conrad}, {Couture}, {Delaney}, {Epps}, {Hilyard}, {Hilyard},
  {Horn}, {Jern}, {Kanto}, {Keane}, {Kibrick}, {Lewis}, {Osborne},
  {Pardeilhan}, {Pfister}, {Ricketts}, {Robinson}, {Stover}, {Tucker}, {Ward},
  \& {Wei}}]{Vogt94}
{Vogt}, S.~S., {et~al.} 1994, in Society of Photo-Optical Instrumentation
  Engineers (SPIE) Conference Series, Vol. 2198, Society of Photo-Optical
  Instrumentation Engineers (SPIE) Conference Series, ed. D.~L. {Crawford} \&
  E.~R. {Craine}, 362

\bibitem[{{Winn} {et~al.}(2010{\natexlab{a}}){Winn}, {Fabrycky}, {Albrecht}, \&
  {Johnson}}]{Winn10b}
{Winn}, J.~N., {Fabrycky}, D., {Albrecht}, S., \& {Johnson}, J.~A.
  2010{\natexlab{a}}, \apjl, 718, L145

\bibitem[{{Winn} {et~al.}(2009){Winn}, {Johnson}, {Albrecht}, {Howard},
  {Marcy}, {Crossfield}, \& {Holman}}]{Winn09}
{Winn}, J.~N., {Johnson}, J.~A., {Albrecht}, S., {Howard}, A.~W., {Marcy},
  G.~W., {Crossfield}, I.~J., \& {Holman}, M.~J. 2009, \apjl, 703, L99

\bibitem[{{Winn} {et~al.}(2010{\natexlab{b}}){Winn}, {Johnson}, {Howard},
  {Marcy}, {Bakos}, {Hartman}, {Torres}, {Albrecht}, \& {Narita}}]{Winn10a}
{Winn}, J.~N., {et~al.} 2010{\natexlab{b}}, \apj, 718, 575

\bibitem[{{Wright} \& {Howard}(2009)}]{Wright09}
{Wright}, J.~T., \& {Howard}, A.~W. 2009, \apjs, 182, 205

\bibitem[{{Zahn}(2008)}]{Zahn08}
{Zahn}, J.-P. 2008, in EAS Publications Series, Vol.~29, EAS Publications
  Series, ed. M.-J. {Goupil} \& J.-P. {Zahn}, 67--90

\end{thebibliography}
\end{document}